# Observation of vortex stripes in UTe$_2$

Y. F. Wang[1,*], H. X. Yao[1,*], T. Winyard[2,*], Christopher Broyles[3], Shannon Gould[3], J. J. Zhu[1], B. K. Xiang[1], Q. S. He[1], K. Y. Liang[1], Z. J. Li[1], P. H. Zhang[1], B. R. Chen[1], K. Z. Yao[1], Q. Z. Zhou[1], D. F. Agterberg[4], E. Babaev[5], S. Ran[3], Y. H. Wang[1,6,†]

1. *State Key Laboratory of Surface Physics and Department of Physics, Fudan University, Shanghai 200433, China*
2. *Maxwell Institute of Mathematical Sciences and School of Mathematics, University of Edinburgh, Edinburgh, EH9 3FD, United Kingdom*
3. *Department of Physics, Washington University in St. Louis, St. Louis, MO 63130, USA*
4. *Department of Physics, University of Wisconsin-Milwaukee, Milwaukee, Wisconsin 53211, USA*
5. *Department of Physics, Royal Institute of Technology, SE-106 91 Stockholm, Sweden*
6. *Shanghai Research Center for Quantum Sciences, Shanghai 201315, China*

\* These authors contributed equally to this work.
† Email address: wangyhv@fudan.edu.cn

**Abstract**

**Quantum vortices are fundamentally important for properties of superconductors. In conventional type-II superconductor they determine the magnetic response of the system and tend to form regular lattices. UTe$_2$ is a recently discovered heavy fermion superconductor exhibiting many anomalous macroscopic behaviors. However, the question whether it has a multicomponent order parameter remains open. Here, we study magnetic properties of UTe$_2$ by employing scanning superconducting quantum interference device microscopy. We find vortex behavior which is very different from that in ordinary superconductors. We imaged vortices generated by cooling in magnetic field applied along different crystalline directions. While a small out-of-plane magnetic field produces typical isolated vortices, higher field generates vortex stripe patterns which evolve with vortex density. The stripes form at different locations and along different directions in the surface plane when the vortices are crystalized along the crystalline *b* or *c* axes. The behavior is reproduced by our simulation based on an anisotropic two-component order parameter. This study shows that UTe$_2$ has a nontrivial disparity of multiple length scales, placing**

**constraints on multicomponent superconductivity. The tendency of vortex stripe formation and their control by external field may be useful in fluxonics applications.**

**Introduction**

Traditionally, most of superconductors were described by a single complex field: superconductivity associated with electron pairing near a single Fermi surface of a crystal [1–4]. Even if there are multiple bands crossing the Fermi level, they are usually considered to be strongly coupled in the superconducting state so that the order parameter is still well described by a single-component theory. However if the coupling between different bands is weak, frustrated or system breaks multiple symmetries, the resulting multicomponent order parameter may lead to many highly unconventional properties on different length scales [1–3,5–7]. The increased number of degrees of freedom allow, under certain conditions [8], more than one coherence length, characterizing the spatial variation of the order parameter. If the London penetration depth, which determines the spatial variation of the magnetic field, lies in between different coherence length, a new type of superconductor occurs, which is coined as type-1.5 [9–11].

Vortex behavior can also be drastically affected by the degrees of freedom of the superconducting order parameter, arising from multiband physics, triplet pairing, or both [8,12,13]. In a single-component type-II superconductor, vortices with the same vorticity repel each other. Consequently, vortices tend to form a hexagonal or Abrikosov lattices in the absence of random pinning. In a multicomponent superconductor, the length scale hierarchy enabled by the type-1.5 superconductivity leads to both attractive and repulsive vortex interactions. So far such states were searched only in nearly-isotropic systems where length-scale hierarchies are direction-independent [14–18]. However, the length scale hierarchy can be much more complicated and direction-dependent when order parameter is anisotropic [19]. The enriched interactions along different directions may generate more exotic vortex structures, such as skyrmion vortices [20]. When multiple components arise from equal-spin pairing, they lead to different magnetic response such as formation of double-quanta vortices and half-quantum vortex stripes [9–11]. Conversely, unconventional vortex patterns in the absence of substantial vortex pinning is a manifestation of the structure of the underlying order parameter.

$UTe_2$ is a recently discovered heavy fermion superconductor which is in the quantum critical regime of incipient magnetism [21,22]. Its Cooper pair spin symmetry is believed to be a triplet [23]. However, the initial evidence for time-reversal symmetry breaking [24] has been controversial [25,26]. The double transition observed in heat capacity as a function of temperature [27] has been attributed to inhomogeneity in the sample [28]. Angle-resolved photoemission spectroscopy [29], density functional theory calculations [30] and quantum oscillation measurements [31–33] all suggest there are multiple bands at the Fermi surface, which could be important for superconductivity. There is also evidence for anisotropic superconductivity: when the magnetic field is along the [011] direction (Fig. 1a), field-reentrant superconductivity was observed above the metamagnetic phase transition [34,35]. As the symmetry of the order parameter is essential for understanding the superconductivity [36], we need suitable experimental techniques to probe $UTe_2$ on the mesoscopic scale in order to avoid sample inhomogeneity.

Detection of unconventional vortex structures require high resolution magnetic microscopy that are directly sensitive to the flux. Generating vortex stripe arrangements along different symmetry directions by applying vectorial magnetic fields is important for understanding the symmetry of the order parameter with respect to the band structure. In this work, we use scanning superconducting quantum interference device (sSQUID) microscopy to perform magnetic imaging of $UTe_2$ under magnetic field applied during cooling through $T_C$. Since vortex striping only appears with enough density of vortices, it is challenging for the nano-SQUID fabricated on a chip to directly distinguish stripes from a large flux background of vortices. To overcome that, we mount the nano-SQUID on a tuning fork [37–40] and the resonant oscillation of the probe allows us to sense minute flux variations with higher sensitivity.

**Inducing vortices in $UTe_2$**
By changing the temperature while measuring susceptibility at a fixed position on the sample, we find that the transition temperature $T_C = 1.57$ K (Fig. 1b). This value is slightly lower than the samples grown by the flux method [41,42], but is very similar to that of the samples grown by the similar method [SOM]. Notably, there is no double-transition from the local susceptometry by our sSQUID (Fig. 1b), which is consistent with the recent finding from the heat capacity measurement that the double-transition disappears at different parts of the crystal. The single $T_C$ of a multi-component order parameter suggests that the components share the same irreducible representation of the symmetry group of the normal state [36].

When we cool the sample under nominally zero magnetic field, we find vortex-antivortex patterns (Fig. 1c), which have also been observed in an earlier work [43].

While it is tempting to argue for its relationship with potential spontaneous breaking of time-reversal symmetry based on such vortices, we cannot rule out that it is due to magnetic impurities in the sample. Increasing the field to $H_z = 5$ Oe, there is a large increase of field gradient as a result of Meissner current but no increase of vortex density (Fig. 1d). This is also true for the vortex patterns obtained after finite field coolings when we change the field below $T_C$ (Fig. S10). Such local magnetic behavior is consistent with volumetric magnetic measurements where total magnetization is dominated by Meissner response in small field [44]. This suggests that UTe$_2$ has a large surface barrier for vortex entrance at low temperature. Therefore, it is necessary to perform field cooling to induce vortices.

**Formation of vortex stripes under $H_z$ and evolution with vortex density**
Knowing the condition required to generate vortices, we report vortex physics based on field-cooling in the rest of this work. Field cooling under $H_z = 0.4$ Oe, we find that vortices congregate into stripe-like patterns (Fig. 1e). Repeated field cooling under the same field shows that even though the overall vortex patterns are similar, the distribution of the stripes is different each time (Figs. 1e-g). This rules out that the observed vortex stripes are due to pinning by line defects or grain boundary [45,46]. After changing the fast and slow scanning axes, we find that the pattern stays the same (Fig. S11). This rules out stripes as scanning artefact. The stripes are not straight or parallel, more similar to those seen on tigers than on bees.

By field cooling under different $H_z$, we investigate the dependence of vortex stripes on vortex density (Fig. 2). Close to zero field, we can see isolated vortices at $H_z = 0.03$ Oe (Fig. 2a) and more occupation of sample by vortices at $H_z = 0.09$ Oe (Fig. 2b). There appear more vortices, which start to coalesce into stripes at $H_z = 0.19$ Oe (Fig. 2c) and $H_z = 0.24$ Oe (Fig. 2d). At $H_z = 0.52$ Oe (Fig. 2e), the image is filled with stripes of different length and randomly connected. It becomes difficult to identify individual vortex. Increasing the field further to 0.74 Oe (Fig. 2f) and 1.09 Oe (Fig. 2g), larger patches of flux start of form, suggesting that vortex stripes merge with increasing vortex density. The stripes are indistinguishable at 2 Oe (Fig. 2h) and a large blob of flux takes over the entire scanned area at 5 Oe (Fig. 2i). The flux is reversed but the evolution of the vortex stripe with vortex density is similar when a negative $H_z$ is applied (Fig. 2j-r). A more complete set of images can be found in Figs. S5-6.

In order to quantify the degree of stripiness of the vortices, we employ an image processing protocol that is commonly adopted for image recognition. From the correlation function (see supplementary materials), we can define vortex anisotropy α as the ratio between the vortex correlation along *x* and that along *y*. Extracting α for the field cool images at different $H_z$, we find that the vortex stripes forms around 0.2 Oe and peaks at 0.5 Oe (Fig. 2s), consistent with qualitative features in the images (Fig. 2a-r). Since $H_z = 4$ Oe roughly converts to a flux density of one flux quantum every (2 μm)², a size that is close to our resolution, the reduction of stripiness above 1 Oe may be attributed to the finite resolution of our probe.

**Vortex stripes along different directions nucleated by tilted fields**

Having shown that vortices nucleated in an out-of-plane field form stripes in UTe₂, we explore the effect of adding a field in the surface plane (011). Without the out-of-plane field, field coolings in an in-plane field $H_x$, which is also along the crystalline $\hat{a}$ (Fig. 1a), only generates vortex and antivortex (Figs. 3a-d). The density of vortex/antivortex increases with $H_x$ and isolated vortices are still distinguishable at 4 Oe (Fig. 3d).

The situation is quite different when both $H_x$ and $H_z$ are applied. We fix $H_z$ at 0.34 Oe and change $H_x$ (Figs. 3e-k). Starting at $H_x = 0$, there are signs of forming vortex stripes under this $H_z$ (Fig. 3e). As the magnitude of $H_x$ increases, the vortex stripes become more visible at $H_x = -0.1$ Oe (Fig. 3g). The stripes are wider at $H_x = -0.4$ Oe (Fig. 3h) and -0.73 Oe (Fig. 3i). They coalesce into larger and longer stripes at $H_x = -1$ Oe (Fig. 3j) and -2 Oe (Fig. 3k). Such connected stripes at ($H_x = -2$ Oe, $H_z = 0.34$ Oe) are distinct from both the vortex/antivortex pattern at ($H_x = -2$ Oe, $H_z = 0$) (Fig. 3c) and the diffusive patches at ($H_x = 0$, $H_z = 2$ Oe) (Fig. 2h). This sequence clearly shows that the vectorial total magnetic field, rather than its out-of-plane projection, is important for the vortex stripes. Vortex stripes due to pancake vortex chains [47] may appear under an in-plane magnetic field (observed in highly layered cuprate superconductors [48]). However, this scenario is clearly inconsistent with our field dependence of the stripes: 1) the presence of stripes under $H_z \neq 0$, $H_x = 0$ (Fig. 2); 2) the absence of stripes under $H_z = 0$, $H_x \neq 0$ (Figs. 3a-d); 3) and the different orientations between the in-plane field and the stripes (Figs. 3e-k).

Now we turn to applying field along the two orthogonal crystalline axes. Field along $\hat{b}$ ($H_b$) and $\hat{c}$ ($H_c$) are realized by combining $H_z$ and $H_x$ with the correct proportion and direction (Fig. 1a). Increasing the field amplitude both makes the vortex denser within the stripe and more stripes occur (Figs. 3l-s), just like in the case for $H_z$ (Fig. 2). Furthermore, changing $H_b$ from 0.2 Oe to 0.66 Oe, we find that the stripes align predominantly along 45° and −17° from the horizontal direction (Figs. 3l-o, pink and green dashed lines). In contrast, the stripes under $H_c$ prefer to align along −45° and 73° (Figs. 3p-s, red and blue dashed lines). By employing the same image processing protocol, we find that 45° from horizontal direction has the largest correlation under $H_b$ (Fig. 3t, pink), whereas -45 degree (Fig. 3t, red) shows the largest correlation under $H_c$. The degree of correlation is similar and does not change much with vortex density. The other direction at 73° is less favored under $H_c$ and the correlation appears to fluctuate with vortex density (Fig. 3t, blue). This quantitative analysis suggests that 45° and -45° are the ground state for the vortex stripes under $H_b$ and $H_c$ respectively, whereas 73° is likely a metastable state under $H_c$.

**Anisotropic two-component order parameter**

In order to understand the vortex stripes and their dependence on an applied magnetic field, we must go beyond a single component order parameter. It is sufficient to consider an anisotropic two-component order parameter, which introduces additional coupled length scales [49] (Fig. 4a). Note that UTe$_2$ is known to be a multi-band superconductor with orthogonal anisotropy [29,30]. If the two bands exhibit different anisotropies, these length scales exhibit different hierarchies in different directions [19]. For example, if in one direction the penetration depth is the intermediary length scale $\xi_I < \lambda < \xi_{II}$ between the two coherence lengths (type-1.5) while in the orthogonal direction it dominates at long range $\xi_I < \xi_{II} < \lambda$ (type-II), then vortex interactions are mediated by core-core attraction and current-current repulsion in orthogonal directions. This anisotropic length scale hierarchy is different from the isotropic type-1.5 regimes discussed in other materials [14–18].

For a quantitative description, we follow [50] and consider the most general anisotropic two-component Ginzburg-Landau model (see End Matter). The anisotropy of the model is given by the anisotropy matrices $Q_{\alpha\beta}$ whose form are derived from the Fermi surfaces of the material. For UTe$_2$, the Fermi surfaces are split into two distinct sets for the electron or hole sections that can be approximated as orthogonal lines in the *a/b* plane, that extend into sheets in the *c* direction [33]. We show that the hierarchy of length scales and hence typology is spatially anisotropic (see End Matter and SOM for detail). In particular, we identify a region of the parameter space that exhibits type-II interactions in one direction while exhibiting type-1.5 interactions in the orthogonal direction. This naturally causes the vortices to cluster along the type-1.5 direction while repelling in the orthogonal type-II direction, hence forming stripes that repel each other. This is the ingredient needed to model the striping observed in UTe$_2$.

Using the parameters suggested by the linearization, we use a Newton flow algorithm to find local minima of the Gibbs free energy. We also find lattice solutions or double periodic solutions to the resulting field equations, by minimizing $G$ with respect to the fields and geometry of a doubly periodic unit cell [SOM]. We can see a clear striping effect (Fig. 4b) for out-of-plane vortices under the anisotropic two-component order parameter considered earlier (Fig. 4a). Since sSQUID measures the magnetic flux along the surface normal ($\hat{z}$) of our UTe$_2$ sample, it is generally different from the local magnetic field of the vortices within the sample when the external field is not applied along $\hat{z}$. Therefore, we have to perform a coordinate transformation in the simulation to directly compare with the experiment. In all three directions of the applied field, there appears striping in the absolute values of the two order parameter components as well as magnetic field along $\hat{z}$ (Fig. 4c-e). In particular, for fields along $H_b$ and $H_c$ (Figs. 4d and e), the directions of the stripes after projecting onto the surface plane (011) are consistent with those from the measurement (Figs. 3l-s).

It is important to note that our effective model is minimal. We design it to capture only the most salient qualitative features of the Fermi surface structure of UTe$_2$. Despite this, the effective model explains and agrees with all the key features of the vortex structure that we experimentally observe. One simple example of the order parameter which agrees with our model and the Fermi surface bands of UTe$_2$ is $\psi_1 = \Delta v_y(k)\hat{x}$, $\psi_2 = \Delta v_x(k)\hat{y}$ (Fig. 4f), where $v_x(k)$ and $v_y(k)$ are the band velocities along $\hat{x}$ and $\hat{y}$, respectively, which ensure that the two gap components are strongly peaked on their respective quasi-1D Fermi surface segments. These two gaps have the same symmetry and hence are Josephson coupled. Furthermore, due to the quasi-1D nature of the bands, this Josephson coupling is weak, and these two gaps naturally have the requisite opposite anisotropies to account for our results. In principle, the quasi-1D band structures naturally allow other similar order parameters with different triplet components and different momentum dependencies that are consistent with our results.

From sSQUID microscopy, we do not observe any evidence for chiral superconductivity under zero field cooling (Fig. 1c). Nevertheless, time-reversal symmetry breaking interaction can be included in the free energy expansion, which will not impact our result. The observation of vortex stripes in UTe$_2$ manifests anisotropic degrees of freedom in the order parameter derived from different Fermi surface bands. The hybridization of the interband phase difference may also lead to novel topological excitations such as fractional vortices and skyrmions [20]. Such excitations cost higher energy and searching for them require further studies and new experimental setups.

**Conclusion**

In conclusion, we use sSQUID microscopy to uncover the vortex stripes in UTe$_2$. Our experiments and simulations are not consistent with pinning-driven structure but agree remarkably well and reveal the multicomponent nature of the order parameter with unconventional hierarchy of the length scales. Hence, we provide evidence that UTe$_2$ has multiple superconductivity components and falls outside of conventional type-I/type-II typology. Further theoretical and experimental studies are required to ascertain the precise form of the underlying multicomponent order parameter. The demonstrated vortex structure formation, and control over these structures can potentially be used in fluxonics (the vortex-based energy-efficient computing schemes).

**Acknowledgement**

We would like to acknowledge support by Shanghai Municipal Science and Technology Major Project (Grant No. 2019SHZDZX01), National Key R&D Program of China (Grant No. 2021YFA1400100) and National Natural Science Foundation of China (Grant No. 12150003). Thomas Winyard would like to thank the School of Mathematics at the University of Edinburgh for funding his postdoctoral position. The work at Washington University is supported by McDonnell International Scholars Academy and the U.S. National Science Foundation (NSF) Division of Materials Research Award DMR-2236528. D.F.A. was supported by Department of Energy, Office of Basic Energy Science, Division of Materials Sciences and Engineering under Award No DE-SC0021971. We would like to thank Martin Speight for useful discussion and helping to originally develop some of the numerical techniques used in the paper.

**Data availability**

The data that support the findings of this work are available from the corresponding authors upon reasonable request.

**References**

[1]  J. Nagamatsu, N. Nakagawa, T. Muranaka, Y. Zenitani, and J. Akimitsu, *Superconductivity at 39 K in Magnesium Diboride*, Nature **410**, 63 (2001).

[2]  D. G. Hinks, H. Claus, and J. D. Jorgensen, *The Complex Nature of Superconductivity in*

*MgB2 as Revealed by the Reduced Total Isotope Effect*, Nature **411**, 457 (2001).

[3] H. J. Choi, D. Roundy, H. Sun, M. L. Cohen, and S. G. Louie, *The Origin of the Anomalous Superconducting Properties of MgB2*, Nature **418**, 758 (2002).

[4] M. Tinkham, *Introduction to Superconductivity* (Courier Corporation, 2004).

[5] G. Blumberg, A. Mialitsin, B. S. Dennis, M. V. Klein, N. D. Zhigadlo, and J. Karpinski, *Observation of Leggett's Collective Mode in a Multiband ${\mathrm{MgB}}_{2}$ Superconductor*, Phys. Rev. Lett. **99**, 227002 (2007).

[6] L. Luo et al., *Quantum Coherence Tomography of Light-Controlled Superconductivity*, Nature Physics **19**, 201 (2023).

[7] V. Grinenko et al., *Superconductivity with Broken Time-Reversal Symmetry inside a Superconducting s-Wave State*, Nat. Phys. **16**, 789 (2020).

[8] M. Silaev and E. Babaev, *Microscopic Theory of Type-1.5 Superconductivity in Multiband Systems*, Phys. Rev. B **84**, 094515 (2011).

[9] M. Ichioka, Y. Matsunaga, and K. Machida, *Magnetization Process in a Chiral p-Wave Superconductor with Multidomains*, Phys. Rev. B **71**, 172510 (2005).

[10] J. A. Sauls and M. Eschrig, *Vortices in Chiral, Spin-Triplet Superconductors and Superfluids*, New Journal of Physics **11**, 075008 (2009).

[11] J. Garaud, E. Babaev, T. A. Bojesen, and A. Sudbø, *Lattices of Double-Quanta Vortices and Chirality Inversion in P_x+ip_y Superconductors*, Phys. Rev. B **94**, 104509 (2016).

[12] E. Babaev and M. Speight, *Semi-Meissner State and Neither Type-I nor Type-II Superconductivity in Multicomponent Superconductors*, Phys. Rev. B **72**, 180502 (2005).

[13] Y. Iguchi, R. A. Shi, K. Kihou, C.-H. Lee, M. Barkman, A. L. Benfenati, V. Grinenko, E. Babaev, and K. A. Moler, *Superconducting Vortices Carrying a Temperature-Dependent Fraction of the Flux Quantum*, Science **380**, 1244 (2023).

[14] V. Moshchalkov, M. Menghini, T. Nishio, Q. H. Chen, A. V. Silhanek, V. H. Dao, L. F. Chibotaru, N. D. Zhigadlo, and J. Karpinski, *Type-1.5 Superconductivity*, Phys. Rev. Lett. **102**, 117001 (2009).

[15] C. W. Hicks, J. R. Kirtley, T. M. Lippman, N. C. Koshnick, M. E. Huber, Y. Maeno, W. M. Yuhasz, M. B. Maple, and K. A. Moler, *Limits on Superconductivity-Related Magnetization in Sr 2 RuO 4 and PrOs 4 Sb 12 from Scanning SQUID Microscopy*, Phys. Rev. B **81**, 214501 (2010).

[16] S. J. Ray, A. S. Gibbs, S. J. Bending, P. J. Curran, E. Babaev, C. Baines, A. P. Mackenzie, and S. L. Lee, *Muon-Spin Rotation Measurements of the Vortex State in \mathrmSr_2\mathrmRu\mathrmO_4: Type-1.5 Superconductivity, Vortex Clustering, and a Crossover from a Triangular to a Square Vortex Lattice*, Phys. Rev. B **89**, 094504 (2014).

[17] P. J. Curran, S. J. Bending, A. S. Gibbs, and A. P. Mackenzie, *The Search for Spontaneous Edge Currents in Sr2RuO4 Mesa Structures with Controlled Geometrical Shapes*, Scientific Reports **13**, 12652 (2023).

[18] P. K. Biswas et al., *Coexistence of Type-I and Type-II Superconductivity Signatures in \mathrmZr\mathrmB_12 Probed by Muon Spin Rotation Measurements*, Phys. Rev. B **102**, 144523 (2020).

[19] T. Winyard, M. Silaev, and E. Babaev, *Hierarchies of Length-Scale Based Typology in Anisotropic $U(1)\phantom{\rule{4pt}{0ex}}s$-Wave Multiband Superconductors*, Phys. Rev. B **99**, 064509 (2019).

[20] T. Winyard, M. Silaev, and E. Babaev, *Skyrmion Formation Due to Unconventional Magnetic Modes in Anisotropic Multiband Superconductors*, Phys. Rev. B **99**, 024501 (2019).

[21] D. Aoki, J.-P. Brison, J. Flouquet, K. Ishida, G. Knebel, Y. Tokunaga, and Y. Yanase, *Unconventional Superconductivity in UTe2*, Journal of Physics: Condensed Matter **34**, 243002 (2022).

[22] S. Ran et al., *Nearly Ferromagnetic Spin-Triplet Superconductivity*, Science **365**, 684 (2019).

[23] L. Jiao, S. Howard, S. Ran, Z. Wang, J. O. Rodriguez, M. Sigrist, Z. Wang, N. P. Butch, and V. Madhavan, *Chiral Superconductivity in Heavy-Fermion Metal UTe2*, Nature **579**, 523 (2020).

[24] D. S. Wei, D. Saykin, O. Y. Miller, S. Ran, S. R. Saha, D. F. Agterberg, J. Schmalian, N. P. Butch, J. Paglione, and A. Kapitulnik, *Interplay between Magnetism and Superconductivity in $UTe_2$*, Phys. Rev. B **105**, 024521 (2022).

[25] S. Sundar, S. Gheidi, K. Akintola, A. M. Côté, S. R. Dunsiger, S. Ran, N. P. Butch, S. R. Saha, J. Paglione, and J. E. Sonier, *Coexistence of Ferromagnetic Fluctuations and Superconductivity in the Actinide Superconductor ${\mathrm{UTe}}_{2}$*, Phys. Rev. B **100**, 140502 (2019).

[26] N. Azari, M. Yakovlev, N. Rye, S. R. Dunsiger, S. Sundar, M. M. Bordelon, S. M. Thomas, J. D. Thompson, P. F. S. Rosa, and J. E. Sonier, *Absence of Spontaneous Magnetic Fields Due to Time-Reversal Symmetry Breaking in Bulk Superconducting ${\mathrm{UTe}}_{2}$*, Phys. Rev. Lett. **131**, 226504 (2023).

[27] I. M. Hayes et al., *Multicomponent Superconducting Order Parameter in UTe2*, Science **373**, 797 (2021).

[28] S. Sundar et al., *Ubiquitous Spin Freezing in the Superconducting State of UTe2*, Communications Physics **6**, 24 (2023).

[29] L. Miao et al., *Low Energy Band Structure and Symmetries of ${\mathrm{UTe}}_{2}$ from Angle-Resolved Photoemission Spectroscopy*, Phys. Rev. Lett. **124**, 076401 (2020).

[30] T. Shishidou, H. G. Suh, P. M. R. Brydon, M. Weinert, and D. F. Agterberg, *Topological Band and Superconductivity in UTe2*, Phys. Rev. B **103**, 104504 (2021).

[31] C. Broyles, Z. Rehfuss, H. Siddiquee, J. A. Zhu, K. Zheng, M. Nikolo, D. Graf, J. Singleton, and S. Ran, *Revealing a 3D Fermi Surface Pocket and Electron-Hole Tunneling in ${\mathrm{UTe}}_{2}$ with Quantum Oscillations*, Phys. Rev. Lett. **131**, 036501 (2023).

[32] D. Aoki, I. Sheikin, A. McCollam, J. Ishizuka, Y. Yanase, G. Lapertot, J. Flouquet, and G. Knebel, *De Haas–van Alphen Oscillations for the Field Along c-Axis in UTe2*, J. Phys. Soc. Jpn. **92**, 065002 (2023).

[33] A. G. Eaton et al., *Quasi-2D Fermi Surface in the Anomalous Superconductor UTe2*, Nature Communications **15**, 223 (2024).

[34] S. Ran et al., *Extreme Magnetic Field-Boosted Superconductivity*, Nature Physics **15**, 1250 (2019).

[35] W. Knafo, M. Nardone, M. Vališka, A. Zitouni, G. Lapertot, D. Aoki, G. Knebel, and D. Braithwaite, *Comparison of Two Superconducting Phases Induced by a Magnetic Field in UTe2*, Communications Physics **4**, 40 (2021).

[36] V. P. Mineev, K. Samokhin, and A. V. Malyavkin, *Introduction to Unconventional Superconductivity*, in (1999).

[37] Y. P. Pan, J. J. Zhu, Y. Feng, Y. S. Lin, H. B. Wang, X. Y. Liu, H. Jin, Z. Wang, L. Chen, and Y. H. Wang, *Improving Spatial Resolution of Scanning SQUID Microscopy with an On-Chip

*Design*, Superconductor Science and Technology **34**, 115011 (2021).

[38] A. Finkler et al., *Nano-Sized SQUID-on-Tip for Scanning Probe Microscopy*, Journal of Physics: Conference Series **400**, 052004 (2012).

[39] D. Jiang et al., *Observation of Robust Edge Superconductivity in Fe(Se,Te) under Strong Magnetic Perturbation*, Science Bulletin **66**, 425 (2021).

[40] Y. P. Pan, S. Y. Wang, X. Y. Liu, Y. S. Lin, L. X. Ma, Y. Feng, Z. Wang, L. Chen, and Y. H. Wang, *3D Nano-Bridge-Based SQUID Susceptometers for Scanning Magnetic Imaging of Quantum Materials*, Nanotechnology **30**, 305303 (2019).

[41] H. Sakai, P. Opletal, Y. Tokiwa, E. Yamamoto, Y. Tokunaga, S. Kambe, and Y. Haga, *Single Crystal Growth of Superconducting $\mathrm{UTe}_{2}$ by Molten Salt Flux Method*, Phys. Rev. Mater. **6**, 073401 (2022).

[42] H. Matsumura et al., *Large Reduction in the A-Axis Knight Shift on UTe2 with Tc = 2.1 K*, J. Phys. Soc. Jpn. **92**, 063701 (2023).

[43] Y. Iguchi, H. Man, S. M. Thomas, F. Ronning, P. F. S. Rosa, and K. A. Moler, *Microscopic Imaging Homogeneous and Single Phase Superfluid Density in UTe 2*, Phys. Rev. Lett. **130**, 196003 (2023).

[44] C. Paulsen, G. Knebel, G. Lapertot, D. Braithwaite, A. Pourret, D. Aoki, F. Hardy, J. Flouquet, and J.-P. Brison, *Anomalous Anisotropy of the Lower Critical Field and Meissner Effect in $\mathrm{UTe}_{2}$*, Phys. Rev. B **103**, L180501 (2021).

[45] J. Gutierrez, B. Raes, A. V. Silhanek, L. J. Li, N. D. Zhigadlo, J. Karpinski, J. Tempere, and V. V. Moshchalkov, *Scanning Hall Probe Microscopy of Unconventional Vortex Patterns in the Two-Gap MgB$_{2}$ Superconductor*, Phys. Rev. B **85**, 094511 (2012).

[46] S. I. Davis, R. R. Ullah, C. Adamo, C. A. Watson, J. R. Kirtley, M. R. Beasley, S. A. Kivelson, and K. A. Moler, *Spatially Modulated Susceptibility in Thin Film $\mathrm{La}_{2-x}\mathrm{Ba}_{x}\mathrm{CuO}_{4}$*, Phys. Rev. B **98**, 014506 (2018).

[47] J. R. Clem, *Two-Dimensional Vortices in a Stack of Thin Superconducting Films: A Model for High-Temperature Superconducting Multilayers*, Phys. Rev. B **43**, 7837 (1991).

[48] A. Grigorenko, S. Bending, T. Tamegai, S. Ooi, and M. Henini, *A One-Dimensional Chain State of Vortex Matter*, Nature **414**, 728 (2001).

[49] M. Silaev, T. Winyard, and E. Babaev, *Non-London Electrodynamics in a Multiband London Model: Anisotropy-Induced Nonlocalities and Multiple Magnetic Field Penetration Lengths*, Phys. Rev. B **97**, 174504 (2018).

[50] M. Speight, T. Winyard, and E. Babaev, *Chiral $p$-Wave Superconductors Have Complex Coherence and Magnetic Field Penetration Lengths*, Phys. Rev. B **100**, 174514 (2019).

[51] M. Silaev and E. Babaev, *Microscopic Derivation of Two-Component Ginzburg-Landau Model and Conditions of Its Applicability in Two-Band Systems*, Phys. Rev. B **85**, 134514 (2012).

[52] I. Timoshuk and E. Babaev, *Microscopic Solutions for Vortex Clustering in Two-Band Type-1.5 Superconductor*, arXiv Preprint arXiv:2404.11491 (2024).

[53] F. Theuss et al., *Single-Component Superconductivity in UTe2 at Ambient Pressure*, Nature Physics (2024).

**End Matter**

The most general anisotropic two-component Ginzburg-Landau model we consider has the Gibbs free energy functional

$$G(\psi_\alpha, A) = \int \left\{ \frac{1}{2} Q_{ij}^{\alpha\beta} \overline{D_i \psi_\alpha} D_j \psi_\beta + \frac{1}{2} |B - H|^2 + F_P(\psi_\alpha) \right\}$$

where $D = d - iA$ is the covariant derivative associated with the $U(1)$ gauge field A. The local magnetic field is then given by $B = \nabla \times A$ and $\Omega$ is the superconducting domain. There are also 2 complex fields $\psi_\alpha = |\psi_\alpha| e^{-i\varphi_\alpha}$ which represent the Cooper pairing in different superconducting bands. Note that Greek indices $\alpha \in [1, 2]$ will always enumerate components of the two order parameters and Latin indices $i = 1, 2, 3$ indicate spatial components, while summation over repeated indices is implied for both. Finally, the constant $H$ is the externally applied magnetic field. $F_P$ collects together the potential terms, which due to gauge invariance, depend only on the condensate magnitudes $|\psi_\alpha|$ and the phase difference between the two condensates $\varphi_{12} = \varphi_1 - \varphi_2$. While in this work we resort to a phenomenological Ginzburg-Landau description, we note that these kinds of expansions were microscopically justified [51] also fully microscopic models, that do not rely on Ginzburg-Landau expansion were shown to exhibit multiple lengths scales despite the presence of interband couplings [8,52].

The anisotropy of the model is given by the anisotropy matrices $Q_{\alpha\beta}$ whose form are derived from the Fermi surfaces of the material. We will use the simplest model that captures this symmetry, which we write as,

$$Q^{11} = \begin{pmatrix} k_{x,1} & 0 & 0 \\ 0 & k_{y,1} & 0 \\ 0 & 0 & k_{z,1} \end{pmatrix}, \quad Q^{22} = \begin{pmatrix} k_{x,2} & 0 & 0 \\ 0 & k_{y,2} & 0 \\ 0 & 0 & k_{z,2} \end{pmatrix}, \quad Q^{12} = 0,$$

where $k_{x,1} \gg k_{y,1}$ and $k_{y,2} \gg k_{x,2}$ to produce the orthogonal anisotropy (note that this anisotropy cannot be removed using spatial rescaling due to the multiple components). In addition, we assume the potential term for the energy functional is:

$$F_P = -\alpha_1 |\psi_1|^2 - \alpha_2 |\psi_2|^2 + \frac{\beta_1}{2} |\psi_1|^4 + \frac{\beta_2}{2} |\psi_2|^4 + \gamma_{12} |\psi_1||\psi_2| \cos \varphi_{12}$$

The final term of the potential term $F_P$ is the Josephson term ($\gamma_{12} < 0$) which breaks the U(1)×U(1) symmetry fixing the unique (up to gauge) state (that minimizes $F_P$) to $(|\psi_\alpha|^0, \varphi_{12}{}^0) = (u_\alpha, 0)$. By linearizing our model about the vacuum state $(|\psi_\alpha|^0, \varphi_{12}{}^0) = (u_\alpha, 0)$, we show that the hierarchy of length scales and hence typology is spatially anisotropic.

This model is discussed in detail in [19,49] where it is shown there are multiple magnetic length scales leading to non-monotonic vortex interactions. Because the model that we study has Josephson coupling between components, sometimes such models are termed single-component [53]. It can be straightforwardly generalized to more general multicomponent case of time reversal symmetry breaking, which is still debated for UTe$_2$ [21,53]. However, even in this simplest non-trivial anisotropic multicomponent model, the anisotropy is sufficient to induce magnetic field inversion [49], fractional vortex splitting [20], leading to dipolar intervortex forces, and directionally dependent typology and interactions leading to striping [19].

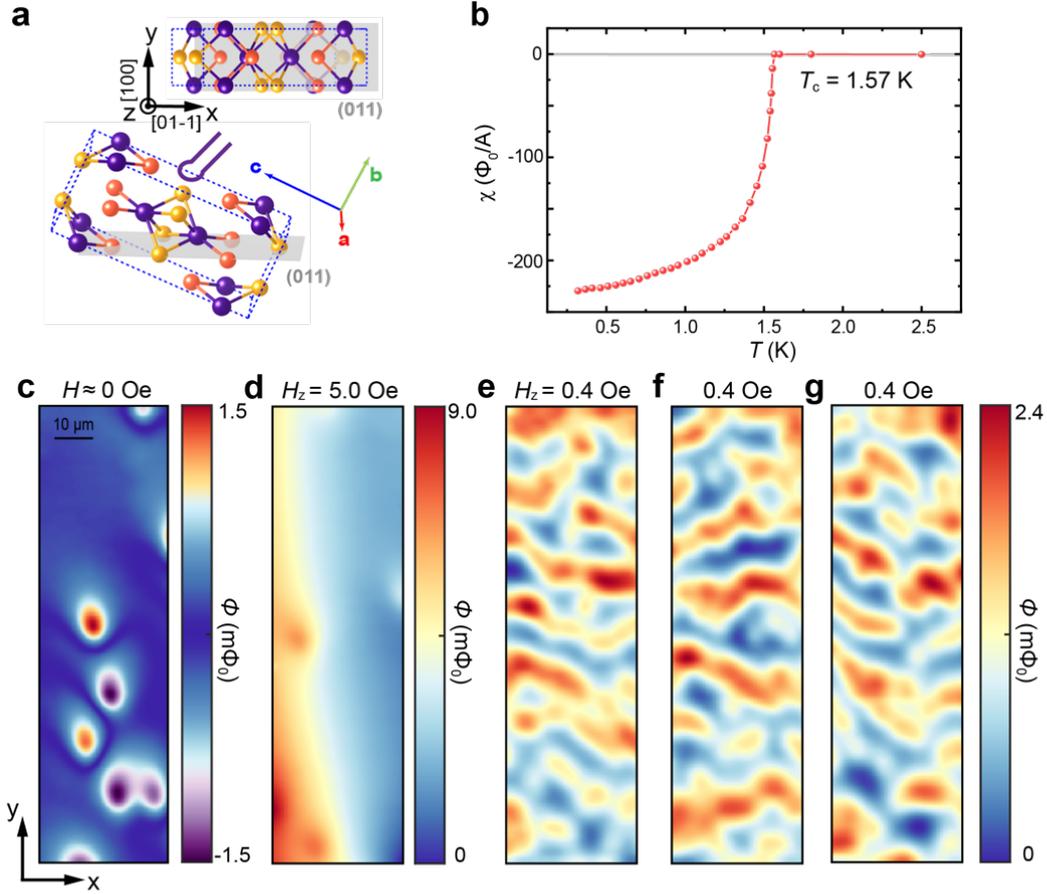

**Figure 1. Vortex structure of UTe₂ after zero-field cooling and field cooling imaged by scanning SQUID. a,** Schematic of the coordinate system of the scanning SQUID microscope with respect to the crystal orientation of UTe$_2$ single crystal. The crystal $\hat{a}$-axis is along the $\hat{x}$ of the microscope, which is the horizontal direction of the images. The crystal naturally exfoliates in the (011) plane, over which our nano-SQUID (purple) scans and picks up flux along $\hat{z}$. **b,** Local diamagnetic susceptibility measured by scanning SQUID as a function of temperature $T$. The critical temperature $T_C = 1.57$ K. **c,** Magnetometry image obtained near the zero field after zero-field cooling showing vortices and anti-vortices. **d,** Magnetometry image obtained after zero field cooling and under an applied out-of-plane field $H_z = 5$ Oe. The field gradient is induced by the Meissner current. **e-g,** Magnetometry images obtained over the same sample area after field cooling under $H_z = 0.4$ Oe, showing striping of vortices. The difference in the stripes from different cycles suggest that they are not induced by vortex trapping. All images are acquired at 500 mK throughout this work.

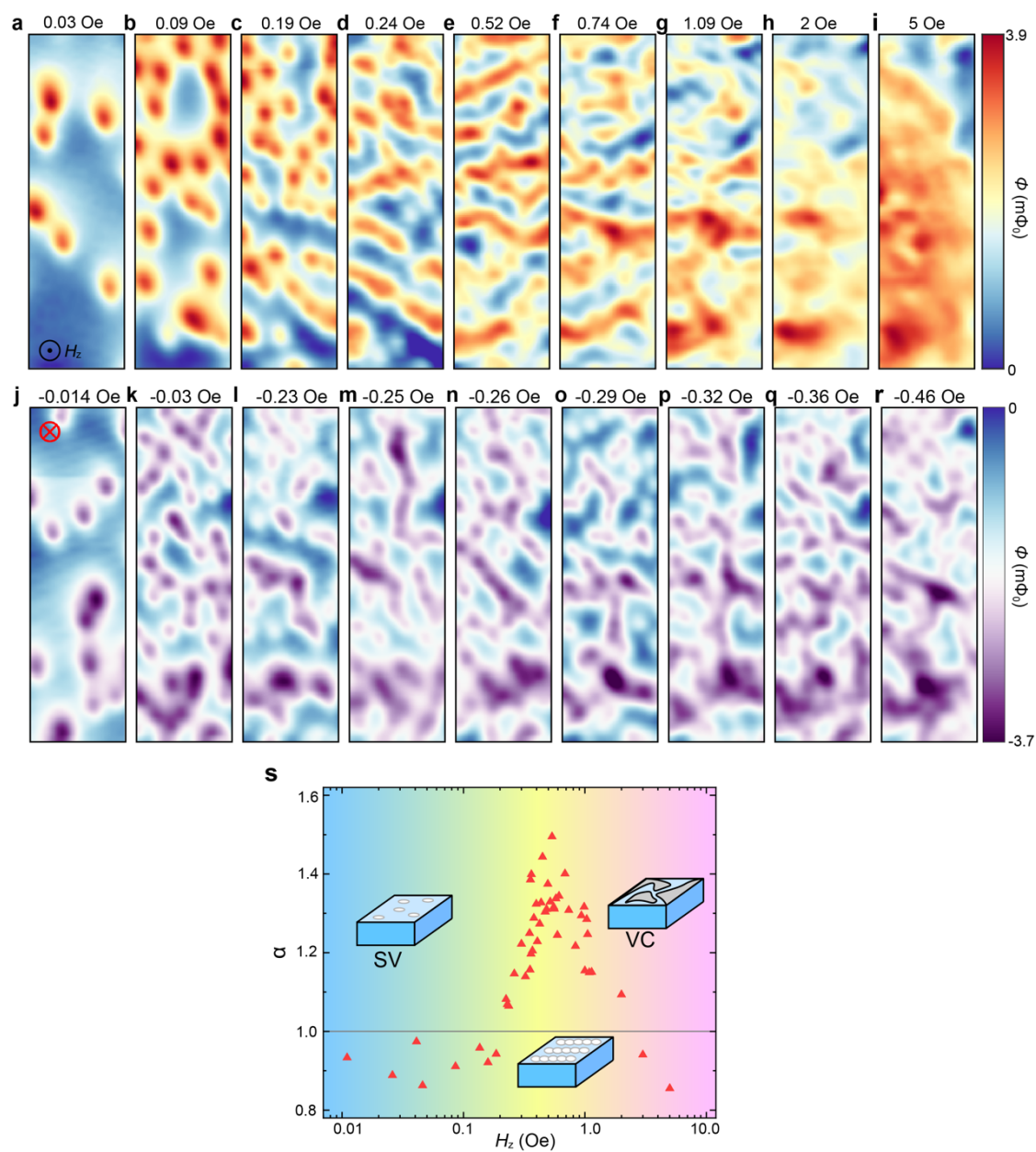

**Figure 2. Evolution of the vortex stripes as a function of flux density. a-r,** Magnetometry images after field cooling under various $H_z$. (More images in different applied fields from the series are available in the supplementary materials.) **a-i** and **j-r** correspond to positive and negative values of $H_z$, respectively. **s,** Vortex anisotropy α extracted from magnetometry images. α is defined as the ratio between the vortex correlation along $\hat{x}$ and that along $\hat{y}$ (see text). $α = 1$ corresponds to patterns without anisotropy in correlations. Vortex stripes are most visible in the density regime where $α > 1$. For the regime of $α < 1$ and $α > 1$, vortices appear as the single vortex (SV) and vortices clusters (VC) in the schematic respectively.

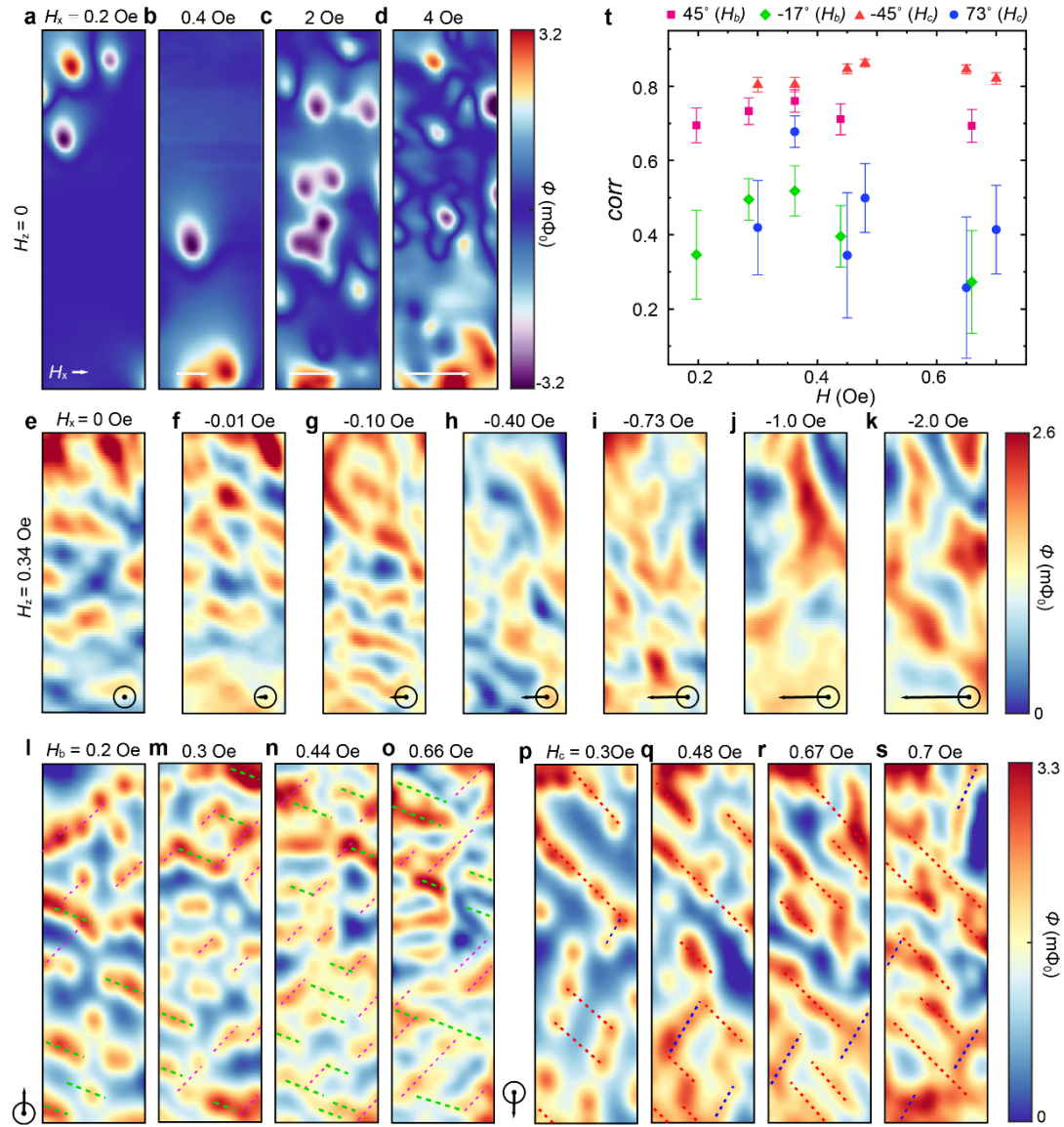

**Figure 3. Stripe patterns after field cooling under tilted field. a-d,** Magnetometry images of vortices after field cooling under field along $H_x$. **e-k,** Magnetometry images after field cooling under a combination of $H_z$ and $H_x$. $H_z$ is fixed at 0.34 Oe while $H_x$ is different for each panel. **l-o** and **p-s,** Magnetometry images after field cooling under field along $\hat{b}$ ($H_b$) and $\hat{c}$ ($H_c$), respectively. The pink and green dashed lines demark the 45° and −17° angles from the horizontal line, respectively; whereas the red and blue dashed lines demark −45° and 73° angles, respectively. **t,** Vortex correlation as a function of field strength. The most favored orientation is 45° for $H_b$ (pink squares) and −45° (red triangles) for $H_c$, respectively. The stripes along −17° under $H_b$ and 73° under $H_c$ appear to be metastable states with weaker correlation that changes with vortex density.

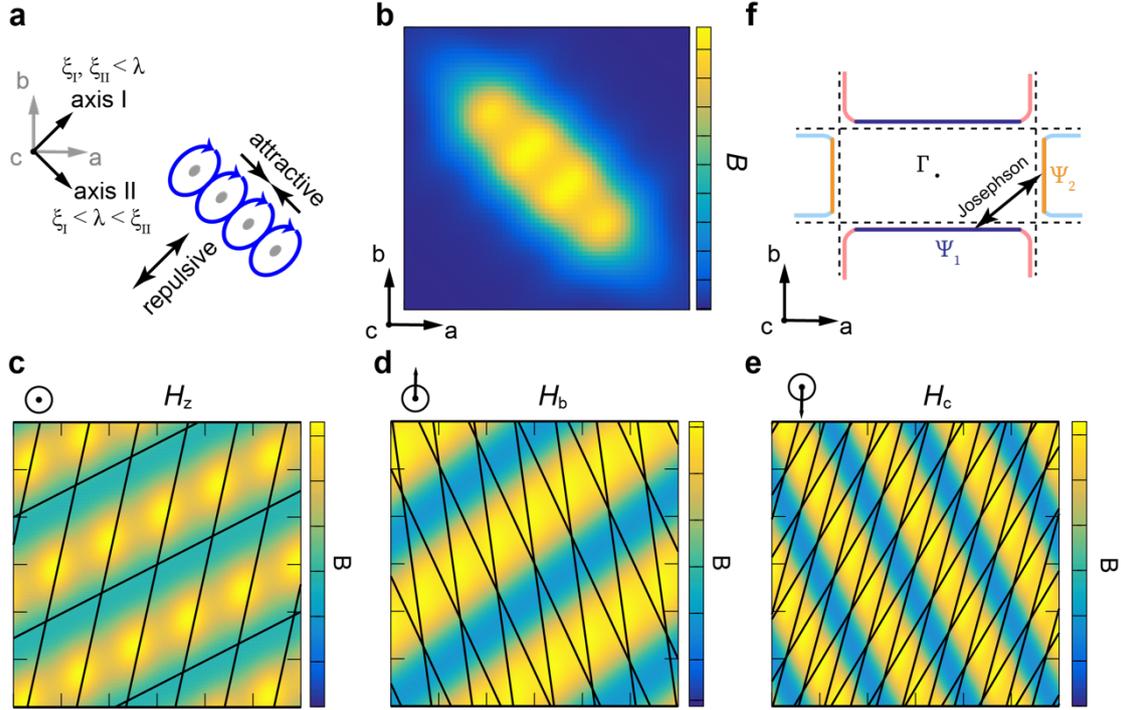

**Figure 4. Simulation of vortex stripes with an anisotropic two-component order parameter. a,** Illustration of the formation of vortex stripes when the vortices are formed under an out-of-plane magnetic field. We consider a two-component order parameter which is highly anisotropic along two in-plane orthogonal axes (axis I and axis II). Along the chains the magnetic field penetration length is the intermediate length scale between two coherence lengths $\xi_I < \lambda < \xi_{II}$ (forming type-1.5 hierarchy). In the direction orthogonal to the chains the magnetic field penetration length scale is the largest length scale. The vortices are repulsive along axis I but attractive along axis II, forming the stripes. (See SOM for more discussions on hierarchy of length scales.) **b,** Plot of the local out-of-plane magnetic field under the model illustrated in **a**. **c-e,** Simulated local magnetic field $B_z$ for the applied field $H$ oriented in $H_z$, $H_b$ and $H_c$. respectively. **f,** Illustration of a slice of the Fermi surface with two bands (blue and pink curves) mainly responsible for $\psi_1$ (purple lines) and $\psi_2$ (orange lines), respectively.

**Contents of the Supplementary Materials**